# Metasurface reconfiguration through lithium ion intercalation in a transition metal oxide

*Simone Zanotto[1], Alessandra Blancato[2], Annika Buchheit[3], Marina Muñoz-Castro[4], Hans-Dieter Wiemhöfer[3], Francesco Morichetti[2], and Andrea Melloni[2]*

[1] Dipartimento di Elettronica, Informazione e Bioingegneria, Politecnico di Milano, Piazza Leonardo da Vinci 32, 20133 Milano, Italy
Present address: Istituto Nanoscienze - CNR, Laboratorio NEST, Scuola Normale Superiore, Piazza San Silvestro 12, 56127 Pisa, Italy
E-mail: simone.zanotto@nano.cnr.it
[2] Dipartimento di Elettronica, Informazione e Bioingegneria, Politecnico di Milano, Piazza Leonardo da Vinci 32, 20133 Milano, Italy
[3] Institute of Inorganic and Analytical Chemistry, University of Münster, Corrensstr. 28/30, 48149 Münster, Germany
[4] Institute of Materials Physics, University of Münster, Wilhelm-Klemm-Straße 10, 48149 Münster, Germany

In the latest years the optical engineer's toolbox has welcomed a new concept, the metasurface. In a metasurface, properly tailored material inclusions are able to reshape the electromagnetic field of an incident beam. Change of amplitude, phase and polarization can be addressed within a thickness of only a fraction of a wavelength. By means of this concept, a radical gain in compactness of optical components is foreseen, even of the most complex ones; [1, 2, 3, 4] other unique features like that of analog computing have also been identified. [5] With this huge potential ready to be disclosed, lack of tunability is still a main barrier to be broken. Metasurfaces must now be made reconfigurable, i.e. able to modify and memorize their state, possibly with a small amount of energy. In this Communication we report low-energy, self-holding metasurface reconfiguration through lithium intercalation in a vanadium pentoxide layer integrated within the photonic device. By a proper meta-atom design, operation on amplitude and phase of linearly polarized light has been demonstrated. In addition, manipulation of circularly polarized light in the form of tunable chirality and tunable handedness-preserving reflection has been implemented. These operations are accomplished using as low as ~50 pJ/$\mu m^2$, raising lithium intercalation in transition metal oxides as one of the most energy efficient self-holding tuning mechanisms known so far for metasurfaces, with significant perspectives in the whole field of nanophotonics.

At the heart of every tunable optical device there is either a moving component or a medium whose optical response functions (permittivity, permeability) change. While the technology of mechanically reconfigurable metasurfaces is generating significant outcomes, [6] there are many applications which would benefit from a tunable device without any moving part. Being permeability in ordinary materials essentially equal to that in vacuum, researchers have intensively looked for processes enabling to tune the permittivity of some of the components which constitute the metasurface. A number of mechanisms and materials have been explored, spanning from free charge in ordinary semiconductors [7, 8] and in graphene [9] up to phase-change materials [10, 11, 12], gas adsorption into metals, [13] and liquid crystals. [14]

An intrinsically different mechanism for tuning the permittivity of a material is electrochromism. In essence, electrochromism consists of a modification of the optical properties of the material occurring when it undergoes an electrochemical reaction. While being well known since many decades, its application to photonic devices is still in its infancy. Indeed, only a few reports can be found, [15, 16] where the active material is a conducting polymer. However, electrochromism occurs also in transition metal oxides, which, compared to polymers, are in general more stable with respect to optical excitation.

Among transition metal oxides, vanadium pentoxide ($V_2O_5$) has the remarkable property to withstand the intercalation of huge amounts of lithium ions in its lattice [17, 18, 19] undergoing an intense investigation from the rechargeable battery community. Concurrently with lithium intercalation, the near-infrared optical properties of $V_2O_5$ are strongly modified, thus enabling the development of compact reconfigurable photonic devices operating at telecommunication wavelengths. As a prototypical device we developed a metasurface comprising an array of aluminum nanoantennas directly placed on top of a $V_2O_5$ layer (**Figure 1 a-b**). The full layer stack (see the Methods section within the Supporting Information for details about the fabrication) also includes a platinum back plane, which has the main function of collecting/injecting the balancing charge carried by electrons during the electrochemical intercalation process. This process took place in an electrochemical cell, following the procedure detailed in the Supporting Information. As an effect of intercalating a molar concentration $x$ of Li ions, the $V_2O_5$ material (which hence becomes $Li_xV_2O_5$) changes its permittivity according to Figure 1 c: the major effect is observed on the imaginary part $\varepsilon''$, with a broad peak extending across the whole near-infrared spectral region, whose values exceed unity. This change of the complex permittivity is related to the mechanism of small polaron hopping, that is to the transfer of conduction electrons bound to the lattice ions to neighbouring orbitals (see Supplementary Information). It should be noticed that in its deintercalated state ($x = 0$) the $V_2O_5$ layer has a small $\varepsilon''$ (about 0.04 at a wavelength of 1550 nm). Together with the large modulation of the absolute value of the complex permittivity $|\varepsilon|$, this implies that vanadium pentoxide deserves significant potentials for both amplitude and phase optical actuators. [20]

In the current geometry, the permittivity of $Li_xV_2O_5$ affects the resonances of the metasurface, whose optical near-field is strongly overlapped with the $V_2O_5$ layer. The localization of the near-field originates from two kinds of resonances. One is the nanoantenna resonance (Figure 1 d shows a single unit cell), i.e., the half-wave dipole resonance (DR) whose field is strongly confined at the metal-dielectric interface. The other is the guided mode resonance (GMR), which can be excited when the periodic pattern allows to couple the free-space radiation into the slab guided mode (Figure 1 e). These two resonances define the metasurface functionality, as will be discussed in the following.

The first functionality we address is the amplitude and phase manipulation of linearly polarized light. In this regards, both metal- and dielectric-based metasurfaces proved to be very effective systems, [21] being also at the core of more complex functions like lensing and beam shaping. [4, 11] We designed the metasurface unit cell with the aid of an optimization algorithm, detailed in the Supporting Information, with the target of maximizing the modulation of linear dichroism around 1550 nm. A centrosymmetric unit cell has been chosen, such that *p*- and *s*-polarized waves do not get mixed by the reflection process, as long as the incident beam lies on a symmetry plane of the structure (see **Figure 2 a**; we choose an incidence angle of 15° since it is the smallest value compatible with the constraints of the measurement setup). According to the optimization procedure, and complying with the fabrication tolerances, the fabricated device has the unit-cell parameters reported in Figure 2 b. The thickness of the $V_2O_5$ layer is 500 nm.

To assess the response of the metasurface to linearly polarized light, we combined reflectometric and ellipsometric analysis (see Supporting Information for technical details). The results of polarization-resolved reflectometry are reported in Figure 2 c. When the Li content in $Li_xV_2O_5$ is zero, the metasurface shows a pronounced linear dichroism around 1500 nm, as highlighted by the dark red band. As the lithium content is increased, the difference between *p*- and *s*-polarized reflectance decreases, and it eventually vanishes for a lithium content $x = 0.18$. It should be noticed that the reflectance averaged over the polarization (i.e., the reflectance of unpolarized light) is mostly unaffected by the lithium content, remaining around the value of 0.25. This behavior is well in agreement with the simulations reported in Figure 2 d, where some additional features are also present, such as the double-dip in the 1500-1600 nm region. These two dips correspond to the aforementioned DR and GMR, whose resonant fields are shown in Figs. 2 d-e, respectively. The coalescence of the double-peaked structure in a single-peaked one can be attributed to the surface roughness of $V_2O_5$ (not included in the electromagnetic simulation), which typically smears sharp resonant features. In contrast, the linear dichroism survives to roughness effects, thus demonstrating the robustness of the metasurface to manipulate this property of light. Yet, roughness depends strongly on the layer thickness thus it can be minimized for thinner samples, and also mitigated by other deposition techniques, [22] which makes the simulated performance practically reachable by actual devices.

As a general consequence of resonance suppression, also the phase of the reflected light is expected to change by increasing losses in the intercalated $Li_xV_2O_5$. Since DR and GMR resonances are polarized, it is expected that intercalation leads to a polarization-dependent phase shift, i.e., to a modulation of the linear birefringence. The effect is mostly relevant in the spectral region between 1600 and 1700 nm, as highlighted by the light red band in Figs. 2 e-f. There, we report the angle $\Delta = \arg(r_p/r_s)$, being $r_{p,s}$ the amplitude reflection coefficients measured with ellipsometry for *p*- and *s*-polarized waves, respectively. The experimentally observed effect is less pronounced than that predicted by simulation, roughness being again the principal cause to be suspected. Nevertheless, the effect is clear, and, as long as the x = 0 and x = 0.03 Li content are considered, phase modulation is not accompanied by a modulation of linear dichroism (see the light red band in Figure 2 b-c). Hence, the metasurface is acting as a tunable birefringent plate, whose thickness is less than the operating wavelength. It should be noticed that better performances can be foreseen, since the present design was not optimized for tunable birefringence operation; rather, this functionality emerged as an effect complementary to linear dichroism. It is however known that proper designs of the nanoantenna geometry and of the layer stack can lead to almost pure phase shift operation with low reflection loss, with potential applications like tunable reflectarrays. [23]

We finally point out that the intercalation process is fully reversible. In Figure 2c and 2e we plotted as red traces the measurements taken after a full deintercalation of the sample, i.e., after having brought the lithium content back from 0.18 to 0. Both polarized reflectance and linear birefringence recovered to their original values in the full spectral range under analysis. This result has a deep technological relevance, since it means that the lithium intercalation/deintercalation process does not damage the aluminum nanoantennas: the basis for $V_2O_5$-based tunable metasurfaces and plasmonic elements is thus firmly grounded.

The second concept we intend to demonstrate is that of a reconfigurable chiral metasurface. Manipulating the chiral states of light is of paramount importance in many fields, like biochemical sensing and telecommunications. Operation with chirality is however weighed down by the fact that most of the photonic components available for this purpose are bulky, since they exploit circular dichroism or birefringence of chiral compounds, which are usually quite small effects. A promising alternative consists instead in relying on metasurfaces, [24] since they offer the advantage of compactness and the opportunity of unique features, like the chiral mirror functionality. [25]

First, we demonstrate that reflection from a $V_2O_5$-based metasurface having a non-centrosymmetric unit cell exhibits wideband, tunable and reversible circular dichroism (CD). We chose one of the simplest unit cell geometry which breaks central symmetry, i.e., the L-shaped antenna. The operating principle of this metasurface is subtle: being the metal pattern single-layered, it must go beyond the Born-Kuhn model for chiral plasmonic dimers. [26] As for the centrosymmetric-cell metasurface analyzed above, the interplay between localized resonances supported by the individual nanoantennas and delocalized resonances originating from the V$_2$O$_5$ slab guided mode dictates the response of the metasurface. The concept of the experiment is schematized in **Figure 3a**: right circularly polarized light (RCP) is in general reflected as both right and left circularly polarized light; this is quantified by the two (intensity) reflectances $R_{RR}$ and $R_{RL}$. Analogously, $R_{LR}$ and $R_{LL}$ can be defined when left circularly polarized (LCP) incident light is considered (not shown). When the metasurface unit cell is non-centrosymmetric (Figure 3b), the response to RCP and LCP can be different, leading to a circular dichroism identified by ($R_{RR}$ + $R_{RL}$) − ($R_{LR}$ + $R_{LL}$). In Figure 3c we plot the wavelength dependent circular dichroism, normalized with respect to the total reflectance $R_{tot}= R_{RR} + R_{RL} + R_{LR} + R_{LL}$ (the normalization procedure is outlined in the Supporting Information). From the measured traces a 50-nm-wide peak of circular dichroism clearly appears around 1520 nm. This peak decreases its intensity as long as lithium is intercalated in V$_2$O$_5$, and eventually gets flattened for a lithium content of 0.18. When lithium is extracted from V$_2$O$_5$, the circular dichroism appears back, as evidenced from the red trace. As it can be observed by comparing Figs. 3c and 3d, the shape and the position of the CD peak is consistent with the result of simulations; again, the small mismatch on the peak value has to the attributed to surface roughness, whose rotational isotropy partially undoes the asymmetry imprinted by the shape of the nanoantennas. It should be noticed that the peak CD pursued in our structure is far larger than that reported in Ref. 27 which, to our knowledge, is the only report of chiral metasurface based on L-shaped monomers. This supports the idea that the interplay between plasmon resonances of the L-shaped particle and delocalized resonances of the underlying dielectric guiding layer are key ingredients for an efficient chiral response, in analogy to what has been observed about the centrosymmetric pattern. A further remark concerns the total reflectance $R_{tot}$, of which we plot in Figure 3g the sole theoretical value due to limitations of the experimental set-up. This quantity remains well different from zero and almost unchanged in the whole range of considered lithium contents. It can hence be argued that the losses induced in the V$_2$O$_5$ layer by the presence of lithium entail the suppression of the resonances responsible for the chiral response of the structure, without having a strong action on the metasurface impedance which remains different from that of air.

The second functionality enabled by the non-centrosymmetric unit cell design is the tuning of the handedness-preserving reflection coefficients. Handedness-preserving reflection is an effect which can be observed in metamaterials where, as opposed to ordinary mirrors, the reflected phase can be opposite for the two orthogonal linear polarizations of incident light. [25, 28] Figure 3e reports the normalized handedness-preserving reflection from the non-centrosymmetric metasurface under analysis: in the region comprised between 1450 and 1700 nm a fraction of reflected light has been detected in the right-to-right and left-to-left channels. By intercalating the V$_2$O$_5$ layer with lithium, the handedness-preserving coefficients vanish, in agreement with what expected from simulations (Figure 3f); again, the observed mismatch, especially on the peak values, should be attributed to roughness. We finally notice that in the spectral region around 1530 nm, highlighted by the light red band, circular dichroism and handedness-preservation coexist: this unique feature, which at its extreme level leads to the concept of *chiral mirror*, may bring to unprecedented applications such as circular-polarization selective Fabry-Pérot laser cavities and ultrasensitive schemes for the detection of chiral molecules. [25] A pure chiral mirror, designed for instance for RCP, fulfils $R_{RR} \neq 0$ and $R_{LL} = R_{RL} = R_{LR} = 0$; in other words, both normalized circular dichroism and normalized handedness-preserving reflection should equal one. To our knowledge, the only geometry which enables pure chiral mirror operation within a subwavelength thickness is that reported in Ref. 25, which employs asymmetric split ring resonators. Unfortunately, it can hardly be applied to devices operating at near-infrared and visible wavelengths, since a simple rescaling of the asymmetric split-ring resonator, originally designed for the GHz spectral range, would require the fabrication of exceedingly small features. Hence, despite the behavior reported in Figure 3 d-f cannot be classified as a pure chiral mirror, we believe that the present concept of L-shaped antennas on a resonant slab could be the beginning of a promising route, which goes beyond the development of $V_2O_5$-based tunable metasurfaces.

Partial and total switching of the metasurface response, i.e., the effect of, respectively, 0.03 and 0.18 lithium intercalation, required only 53 pJ/μm$^2$ and 260 pJ/μm$^2$ (see Supporting Information). These amounts could be further reduced by decreasing the volume of $V_2O_5$, relying, for instance, in ultrathin geometries like those involving gap surface plasmons. [21] Decreasing the volume, along with nanostructuration of $V_2O_5$, [29] is expected to be beneficial also on the switching time. While it was not the main aim of the present work to assess the response time of the $V_2O_5$-based metasurface, and long intercalation times were employed in order to avoid any possible damage of the structure, a reasonable estimate for the typical response time of devices like ours is of the order of a fraction of a second, [29] which fits the requirements of applications like optical bench reconfiguration. In view of this and other applications it would be strongly beneficial to make the reconfiguration process through solid-state electrolytes, which are indeed under active investigation for their significance in the field of battery technology. [30, 31]

In conclusion, we realized a tunable metasurface based on lithium intercalation in vanadium pentoxide. Harnessing the huge modification of its permittivity, we demonstrated an active control over several effects having relevance for polarization-handling optical components of subwavelength thickness. In detail, linear and circular dichroism, linear birefringence, and handedness-preserving reflectance were successfully tuned to wide extents. These operations are pursued in a self-holding, reversible, and energy-saving fashion, reflecting several desirable properties which are intrinsic to the material. We thus believe that the potential applications of vanadium pentoxide and other mixed ionic-electronic conductors (such as $WO_3$) go well further from the field of metasurfaces, opening up interesting technological advances in the whole field of photonics.

**Supporting Information**
See below

**Author contributions:** S. Z. conceived the experiment, designed the sample, performed the nanofabrication and analyzed the data. S. Z., A. Bl., and A. Bu. performed the optical measurements. M. M.-C. prepared the electrochromic oxide film and performed the electrochemical processes. H.-D. W., F. M., and A. M. supervised the study. All authors discussed the results and contributed to the manuscript preparation.

**Acknowledgements**
The authors thank Claudio Somaschini and Giosuè Iseni in staff of Polifab at Politecnico di Milano for the support in the fabrication of the devices, Marcus Bernemann for performing AFM measurements, Amos Egel for the precious suggestions about electromagnetic simulation and D. S. Wiersma for useful discussions. The work was supported by the European Commission through the BBOI project, FP7 323734.

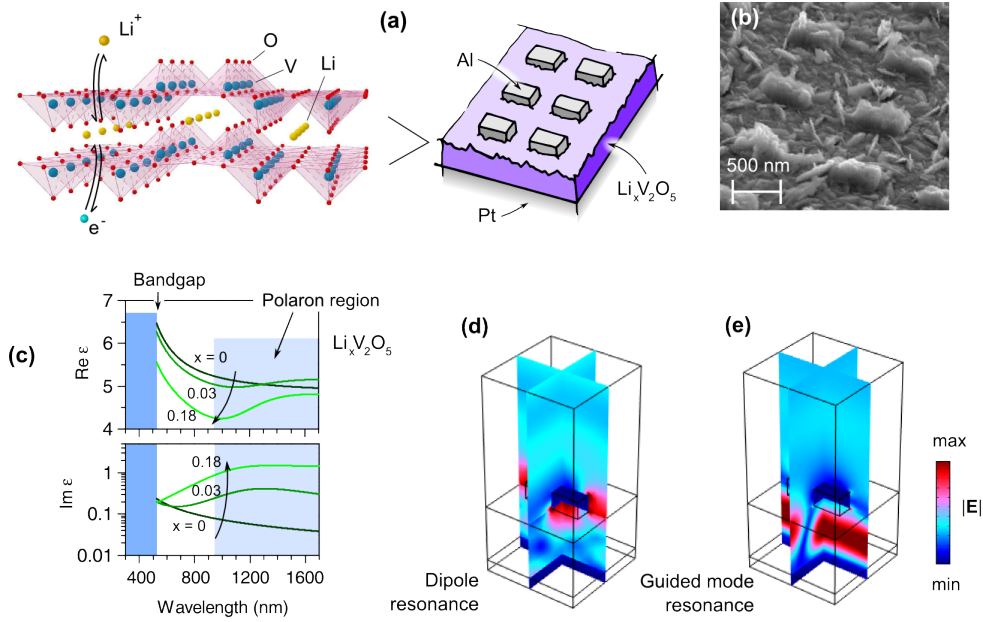

**Figure 1.** Concept of the vanadium pentoxide assisted reconfigurable metasurface. Aluminum nanoantennas are patterned on top of a $V_2O_5$ film (a). A scanning electron micrograph of the fabricated sample is reported in (b). Lithium intercalation into the $V_2O_5$ lattice strongly modifies its permittivity, in both real and imaginary parts (c). Radical effects on the response of the resonant metasurface are thus expected, especially in correspondence to the wavelengths where optical resonances occur (d-e).

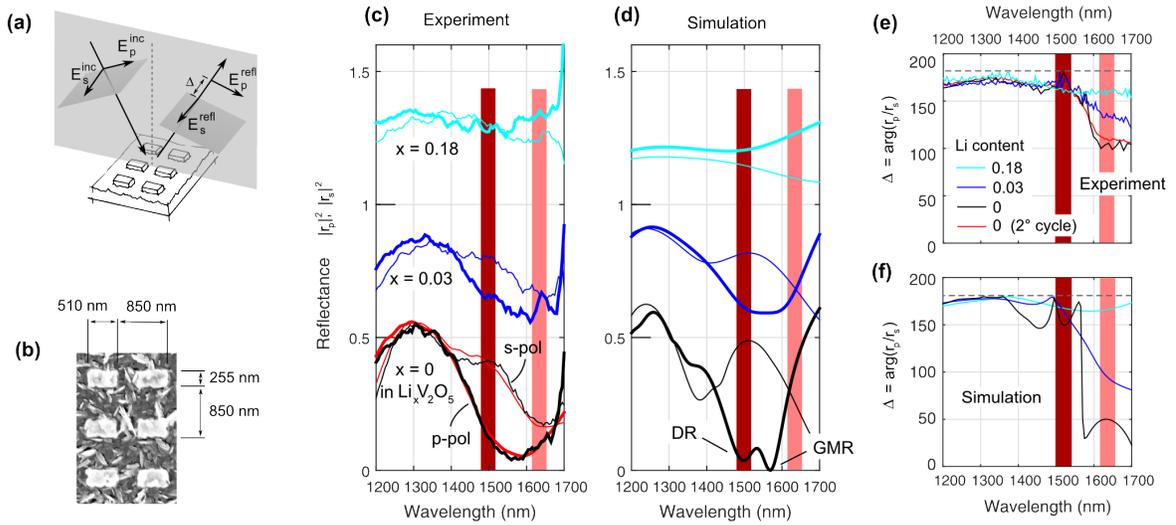

**Figure 2.** Manipulation of linearly polarized waves by means of a centrosymmetric $V_2O_5$-loaded metasurface. Schematic of the experimental framework (a), where linearly polarized plane waves probe the metasurface. Thanks to the pattern symmetry, there is no polarization mixing (i.e., a *p*-wave is reflected into a *p*-wave and similarly occurs for *s*-waves); however, both amplitudes and phases of the reflected fields are affected. The reflection coefficient for *p*-waves is defined as $r_p = E^{refl}_p / E^{inc}_p$, and similarly for *s*.

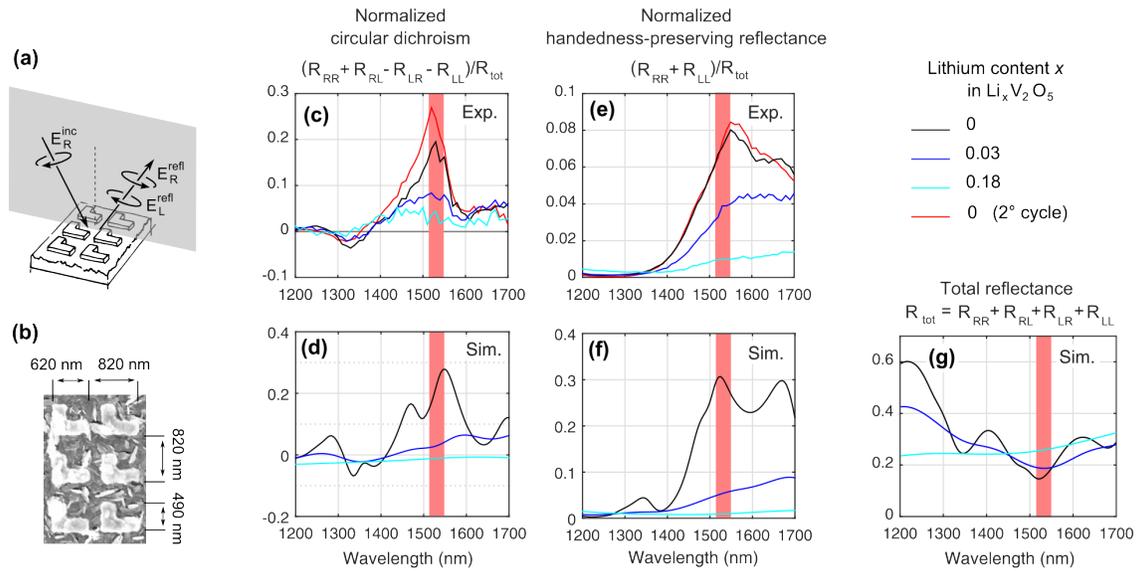

**Figure 3.** Manipulation of circularly polarized waves by means of a non-centrosymmetric $V_2O_5$-loaded metasurface. Schematic of the operation concept (a), where a circularly polarized plane wave probes the metasurface. In the actual experiment the intensity reflection coefficients $R_{RR}$, $R_{RL}$, $R_{LR}$, $R_{LL}$ are obtained indirectly from an ellipsometric measurement (the procedure is detailed in the Supporting Information). Details on the nanoantenna size are given in panel (b). Lacking a center of inversion, the unit cell defines a sense of twist for the metasurface, which is allowed to respond differently to waves of opposite handedness. Circular dichroism is indeed observed (panels c-d) mostly at telecommunication wavelengths. Moreover, as opposed to conventional mirrors, the metasurface exhibits handedness-preserving reflection (panels e-f). In the spectral range highlighted by the light red band this effect coexists with circular dichroism, enabling chiral mirror operation. Total reflectance has instead a weaker dependence upon lithium intercalation (panel g).

# Supporting Information

1. Methods

***Sample fabrication*** The substrate for sample preparation consists of a polished silicon (111) wafer with a 100 nm sputtered Pt layer. The $V_2O_5$ layer was then realized through DC magnetron sputter deposition. The sputter chamber was evacuated to a base pressure of $10^{-8}$ mbar and the target (Vanadium, 99.9% purity) was pre-sputtered in Ar for 60 s to get rid of possible impurities on the target surface. Subsequently, the sputtering gas was fixed to a mixture of $Ar/O_2$ with 40% oxygen content, the working pressure was set to $5 \cdot 10^{-3}$ mbar and a DC sputtering power of 4 W·cm$^{-2}$ was applied to obtain a $V_2O_5$ layer with a thickness gradient centered around 520 nm. Afterwards, the aluminum pattern was realized by means of electron beam lithography followed by physical vapor deposition and lift-off. In detail, a PMMA resist bilayer consisting of AR-P 641-035 and AR-P 679-04 by All Resist was employed. Before resist deposition, the sample was cleaned in acetone and isopropanol (IPA), and baked for 3 min at 150°. Then, the PMMA layers were spun in sequence at 4000 rpm for 1 min each, with a 5 min baking at 150° at the end. Resist was then exposed at 20 keV, 230 µC/cm$^2$, and developed in MIBK:IPA 1:3 for 2 min. This recipe allows for an overall resist thickness of about 500 nm, sufficient enough for planarizing the sharp nanocrystallites of the $V_2O_5$ film, and for a good undercut which resulted in an easy lift-off (warm acetone and gentle solvent squeezing was employed). Centrosymmetric and non-centrosymmetric patterns were covering an area of 400 x 400 µm$^2$ each.

***Electrochemical intercalation/deintercalation*** Electrochemical intercalation and deintercalation of lithium in the $V_2O_5$ thin layer was performed through chronopotentiometry. A three-electrode cell setup was used with the $V_2O_5$ sample as working electrode, where the Pt layer underneath served as current collector. Elemental lithium was used as counter and reference electrode, thus all potentials in this paper are given versus Li/Li$^+$. As liquid electrolyte served a 1 M solution of $LiClO_4$ in a 1:1 mixture of ethylene carbonate and dimethylcarbonate. To prevent undesired reactions of lithium with moisture, the cell was assembled in an argon filled glovebox. The electrochemical experiments were carried out with a VSP-300 multichannel potentiostat/galvanostat (BioLogic), whereas the applied constant current was ±0.2 µA for all the intercalation/deintercalation steps.

***Numerical analysis and sample design*** The electromagnetic response of the metasurface has been studied by rigorous coupled-wave analysis (RCWA), through a custom code which implements the formalism reported in [32, 33, 34]. MATLAB language was used, which enabled to include the calculation routines in an optimization script. Optimization was run for the design of the centrosymmetric structure, by looking for the nanoantenna shape and $V_2O_5$ thickness which maximizes the linear dichroism at 1550 nm. Bounds on the unit cell size have been set in order to avoid diffraction, and the thickness of $V_2O_5$ was constrained to lie between the limits dictated by the prepared film. For what concerns the non-centrosymmetric metasurface, instead, manual inspection led to the design of Figure 3b of the main text. The $V_2O_5$ thickness was there 480 nm. A square truncation scheme with 21 x 21 spatial harmonics has been employed. The outcomes of RCWA for the centrosymmetric metasurface have been cross-checked with those of finite-element modeling (FEM), commercially available from COMSOL. The field maps reported in Figure 1 have been obtained with FEM.

***Optical characterization*** The centrosymmetric pattern has been characterized by both polarized reflectometry and spectroscopic ellipsometry performed with a VASE machine (Woollam Inc.). The non-centrosymmetric pattern, which mixes the polarization of linearly-polarized incident waves, was instead studied through ellipsometry with the auto-retarder option, which outputs the cross-polarization spectroscopic parameters $\Psi_1 = \arctan(|r_{pp}/r_{ss}|)$, $\Psi_2 = \arctan(r_{ps}/r_{pp})$, $\Psi_3 = \arctan(r_{sp}/r_{ss})$, $\Delta_1 = \arg(r_{pp}/r_{ss})$, $\Delta_2 = \arg(r_{ps}/r_{pp})$, $\Delta_3 = \arg(r_{sp}/r_{ss})$. From those parameters, and relying on the relations

$$r_{RR/LL} = (r_{pp} - r_{ss} \mp i(r_{ps} + r_{sp}))/2$$
$$r_{RL/LR} = (r_{pp} + r_{ss} \pm i(r_{ps} - r_{sp}))/2$$

the normalized circular dichroism and the handedness-preserving reflection reported in Figure 3 of the main text were obtained. Since the probe beam spot was larger than the patterns, it was spatially filtered by means of a perforated glossy mask placed in contact with the sample.

## 2. Near-IR behavior of Li-intercalated $V_2O_5$

In $V_2O_5$, as well as in other transition metal oxides (such as $WO_3$, …), conduction electrons do not behave as free charges due to their strong interaction with lattice ions, leading to the creation of polaron particles. Polarons are inherently associated with defects in the crystalline material structure. In an ideal $V_2O_5$ crystal, all the vanadium atoms should be in a nominal $V^{5+}$ valence state, but thermal agitation and crystal defects (e.g. oxygen vacancies) create a small concentration of $V^{4+}$ atoms [35]. Electronic conduction occurs by means of electron hopping, which is equivalent to the motion of these +4 valent vanadium states. The combined transfer of an electron and its accompanying lattice distortion, also referred to as small polaron transport, occurs with a greatly reduced mobility compared to conduction in free-charge transport in covalent semiconductors, such as silicon [36].

Small polarons have strong impact on the optical response of $V_2O_5$, especially in the near infrared range. Near infrared absorption in $V_2O_5$ has been demonstrated to be associated with charge transfer to neighbouring orbitals, that is with polarons which move from $V^{4+}$ to $V^{5+}$ sites by absorbing photons [37, 38], that is

$$h\nu + V_i^{5+} + V_j^{4+} \rightarrow V_i^{4+} + V_j^{5+}$$

where $h\nu$ is the energy of the absorbed photon. The peak of the polaron absorption resulting from the electron transition from $V^{4+}$ sites to $V^{5+}$ sites ranges from the visible region to the near-infrared region [39]. The absorption is broadened by lattice disorder that makes electronic transitions take place between neighbouring sites with a substantial spread in energy. Near-infrared absorption due to small polaron effects in $V_2O_5$ has been observed in films made by evaporation [40], sputtering [41] and sol-gel deposition [42].

In non-intercalated $V_2O_5$ this absorption is related to oxygen vacancies in the lattice [38]: empty 3d orbitals of V atoms adjacent to such vacancies are able to localize excess electrons, which produces $V^{4+}$ pairs in the vicinity of the vacancy. Upon (lithium) intercalation, electrons and ions intercalated into the $V_2O_5$ matrix lead to a change in the valence state of vanadium [43, 44]

$$V^{5+} + e \rightarrow V^{4+}$$

increasing the concentration of $V^{4+}$ in the lithiated $V_2O_5$ films. Thus the near IR absorption increases with intercalation.

In the literature, the optical properties of (lithiated) $V_2O_5$ films has been modeled through a multi-Lorentzian dielectric function [45]. We elaborated this concept performing a similar analysis on an unpatterned region of the film employed for the main experiment, focusing on the spectral range between 550 and 1700 nm. For what concerns the non-lithiated film, a single Lorentz oscillator (modeling the interband transition) was sufficient for a satisfactory description of the optical spectra. When the lithiated case $Li_{0.03}V_2O_5$ is instead considered, a second Lorentz oscillator, modeling the small polaron transition, was needed. Finally, when considering a more lithiated state ($Li_{0.18}V_2O_5$), a third contribution in the form of a Drude term was required to properly describe the optical function. We interpret the need for a Drude term as the presence of free charge, as a consequence of the increased molar fraction of lithium.

In summary, the extrapolated $Li_xV_2O_5$ dielectric function has the following expression

$$\varepsilon_{LixV2O5} = \frac{\omega_{p,ib}^2}{\omega_{0,ib}^2 - \omega^2 - 2i\omega\Gamma_{ib}} + \frac{\omega_{p,pol}^2}{\omega_{0,pol}^2 - \omega^2 - 2i\omega\Gamma_{pol}} + \frac{\omega_{p,D}^2}{-\omega^2 - 2i\omega\Gamma_D} + 1$$

(S1)

where the label *ib* refers to the interband transition, *pol* to the polaron, and *D* to Drude. The numerical values of the parameters are reported in Table I. To better describe the surface roughness, the top 70 nm of the $V_2O_5$ film are treated with the Bruggeman effective medium approximation (EMA) [46]:

$$\varepsilon_{EMA} = \frac{\varepsilon_{LixV2O5} + 1}{2}.$$

The remaining 460 nm (thicknesses have been measured through an AFM) are instead regarded as bulk $Li_xV_2O_5$. Figure S1 reports the measured reflectance spectra and the fitting curves; the fitting parameters are reported in Table I. The permittivity resulting from Equation S1 is reported in the main article, Figure 1c. As it can be noticed from Figure S1, there is a slight mismatch between experiment and theory in the region of the polaron resonance. We attempted to solve this discrepancy by fitting with respect to the thickness of the EMA, and employing different averaging approaches (Lorentz-Lorenz, Maxwell-Garnett), without however getting significant improvements. It is thus likely that more refined models which go beyond the Lorentz-Drude oscillator are required to fully describe the response of lithiated $V_2O_5$. Another possible reason for the mismatch is that the model assumes a uniform distribution of the lithium concentration across the whole $V_2O_5$ thickness; instead, a non-flat distribution could have been present.

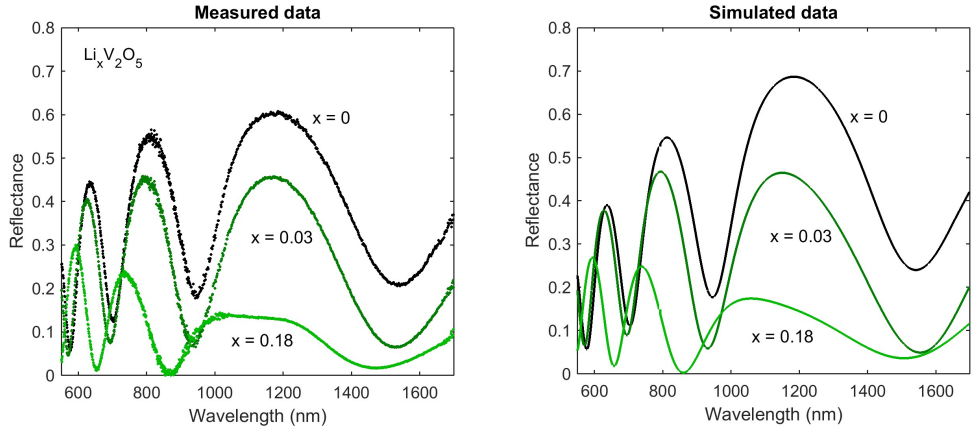

**Figure S1.** Reflectance of an unpatterned $V_2O_5$ film at different intercalation levels.

| Lithium content x | $\hbar\omega_{0,ib}$ | $\hbar\Gamma_{ib}$ | $\hbar\omega_{p,ib}$ | $\hbar\omega_{0,pol}$ | $\hbar\Gamma_{pol}$ | $\hbar\omega_{p,pol}$ | $\hbar\Gamma_D$ | $\hbar\omega_{p,D}$ |
|---|---|---|---|---|---|---|---|---|
| 0 | 4.32 | 0.12 | 8.5 | --- | --- | --- | --- | --- |
| 0.03 | 4.41 | 0.10 | 8.6 | 1.01 | 0.37 | 0.52 | --- | --- |
| 0.18 | 4.60 | 0.03 | 8.7 | 1.07 | 0.36 | 0.81 | 0.72 | 1.11 |

**Table I.** Parameters of the Lorentz-Drude oscillator model for lithiated $V_2O_5$. All quantities are stated in eV.

## 3. Energy required for the actuation of the metasurface

As for every tunable optical device, energy and time required for switching are key metrics determining the performance and the potential applications. In a tunable photonic device based on electrochromism, switching energy and time can be directly deduced from the voltammetric data recorded during the intercalation/deintercalation process.

Figure S2 reports the chronopotentiometry traces recorded throughout the multistep intercalation/deintercalation sequence used to bring the metasurface sample in the lithiation states considered for the optical measurements reported in the main text. A constant current of ±0.2 µA was forced through the metasurface, which acted as the working electrode of the electrochemical cell (see also the Method paragraph). By simply integrating the current-voltage product, the overall energy required to switch the metasurface was obtained; dividing for the sample surface (0.88 cm$^2$) yielded the energy required per unit area. For steps 1+2, i.e, for a 0.03 intercalation which implies a partial switching of the metasurface resonances (see Figures 2 and 3 of the main text), an energy of 53 pJ/µm$^2$ is required. When instead looking for a complete suppression of the metasurface resonances, i.e., when looking for the ≈0.18 lithium intercalation, an energy of 260 pJ/µm$^2$ is needed.

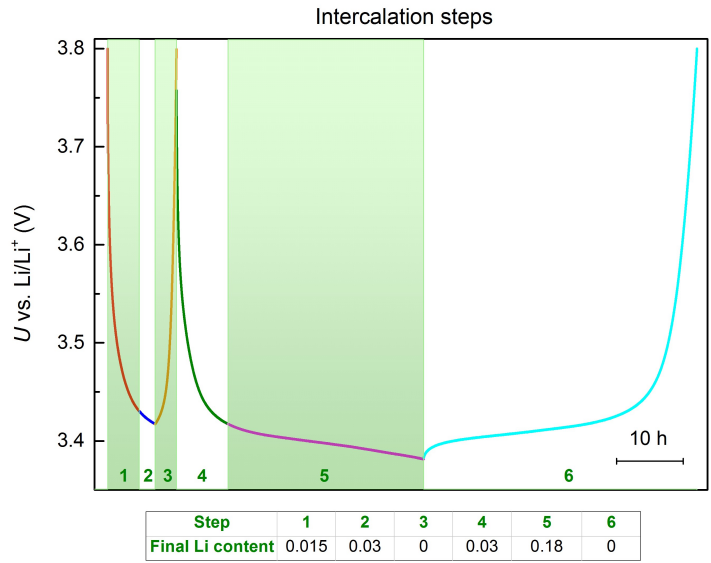

**Figure S2.** Chronopotentiometry of the intercalation/deintercalation process.